%% file: draft.tex
\documentclass{SciPost}

\hypersetup{
    colorlinks,
    linkcolor={red!50!black},
    citecolor={blue!50!black},
    urlcolor={blue!80!black}
}

\usepackage[bitstream-charter]{mathdesign}
\DeclareSymbolFont{usualmathcal}{OMS}{cmsy}{m}{n}
\DeclareSymbolFontAlphabet{\mathcal}{usualmathcal}

\fancypagestyle{SPstyle}{
\fancyhf{}

\fancyfoot[C]{\textbf{\thepage}}
}

\usepackage[T1]{fontenc}

\usepackage{orcidlink}

\usepackage{booktabs}
\usepackage{multirow}
\usepackage{tabularx}

\usepackage[nameinlink]{cleveref}

\usepackage{csquotes}
\usepackage{textcomp}

\begin{document}

\pagestyle{SPstyle}

\begin{center}{\Large \textbf{\color{scipostdeepblue}{
TrackFormers Part 2: Enhanced Transformer-Based Models for High-Energy Physics Track Reconstruction\\
}}}\end{center}

\begin{center}\textbf{
Sascha {Caron}\textsuperscript{1,2}\orcidlink{0000-0003-2941-2829},
Nadezhda {Dobreva}\textsuperscript{1,2}\orcidlink{0009-0006-0923-6054},
Maarten {Kimpel}\textsuperscript{3},
Uraz {Odyurt}\textsuperscript{4}\orcidlink{0000-0003-1094-0234},
Slav {Pshenov}\textsuperscript{2,5}\orcidlink{0009-0002-6971-5269},
Roberto {Ruiz de Austri Bazan}\textsuperscript{6}\orcidlink{0000-0003-3688-9609},
Eugene {Shalugin}\textsuperscript{1,2},
Zef {Wolffs}\textsuperscript{2,5}\orcidlink{0000-0001-5100-2522} and
Yue {Zhao}\textsuperscript{7$\star$}\orcidlink{0009-0005-4074-5116}
}\end{center}

\begin{center}
{\bf 1} High-Energy Physics, Radboud University, Nijmegen, The Netherlands
\\
{\bf 2} National Institute for Subatomic Physics (Nikhef), Amsterdam, The Netherlands
\\
{\bf 3} Institute for Computing and Information Sciences, Radboud University, Nijmegen, The Netherlands
\\
{\bf 4} Faculty of Engineering Technology, University of Twente, Enschede, The Netherlands
\\
{\bf 5} Institute of Physics, University of Amsterdam, Amsterdam, The Netherlands
\\
{\bf 6} Instituto de Física Corpuscular, IFIC-UV/CSIC, Valencia, Spain
\\
{\bf 7} High Performance Machine Learning, SURF, Amsterdam, The Netherlands
\\[\baselineskip]
$\star$ \href{mailto:yue.zhao@surf.nl}{\small yue.zhao@surf.nl}\,
\end{center}

\definecolor{palegray}{gray}{0.95}
\begin{center}
\colorbox{palegray}{
  \begin{tabular}{rr}
  \begin{minipage}{0.37\textwidth}
    \includegraphics[width=60mm]{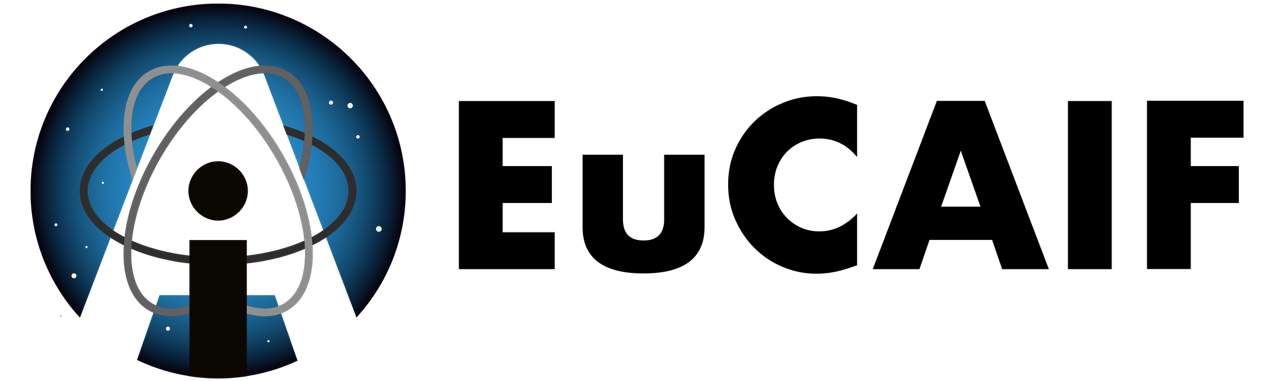}
  \end{minipage}
  &
  \begin{minipage}{0.5\textwidth}
    \vspace{5pt}
    \vspace{0.5\baselineskip} 
    \begin{center} \hspace{5pt}
    {\it The 2nd European AI for Fundamental \\Physics Conference (EuCAIFCon2025)} \\
    {\it Cagliari, Sardinia, 16-20 June 2025
    }
    \vspace{0.5\baselineskip} 
    \vspace{5pt}
    \end{center}
    
  \end{minipage}
\end{tabular}
}
\end{center}

\section*{\color{scipostdeepblue}{Abstract}}
\textbf{\boldmath{%
High-Energy Physics experiments are rapidly escalating in generated data volume, a trend that will intensify with the upcoming High-Luminosity LHC upgrade. This surge in data necessitates critical revisions across the data processing pipeline, with particle track reconstruction being a prime candidate for improvement. In our previous work, we introduced \enquote{TrackFormers}, a collection of Transformer-based one-shot encoder-only models that effectively associate hits with expected tracks. In this study, we extend our earlier efforts by conducting detailed investigations into more custom Transformer attention mechanisms, a new design combining geometric projection and lightweight clustering, and a joint model conditioning classification on a regressor's predictions. Furthermore, we discuss new datasets that allow the training on hit level for a range of physics processes. These developments collectively aim to boost both the accuracy and potentially the efficiency of our tracking models, offering a robust solution to meet the demands of next-generation high-energy physics experiments.
}}

\vspace{\baselineskip}




\input{text/body}

\bibliography{bibliography/references.bib}

\end{document}

%% file: text/body.tex

\section{Introduction}
\label{sec:introduction}
The High-Luminosity LHC (HL-LHC) will generate unprecedented volumes of collision data, creating significant challenges for particle track reconstruction, where hundreds of thousands of detector hits must be accurately associated with their originating particles. Traditional reconstruction methods, while precise, struggle to scale efficiently to these data rates. Transformer-based machine learning models offer a promising alternative: in prior work, we introduced \enquote{TrackFormers}, encoder-only, one-shot transformers that map hits directly to particle tracks. In this study, we extend this approach by exploring a new design combining geometric projection and lightweight clustering, a joint model conditioning classification on a regressor's predictions, and FlexAttention~\cite{dong_flex_2024}. To support future model training and evaluation, we provide a fully reproducible ACTS-based hit-level dataset spanning signal and background processes across multiple pileup levels.

\section{Improved methods}
\label{sec:improved_methods}

\subsection{New datasets}
\label{sec:new_datasets}
We created a new hit-level dataset with a reproducible ACTS-based  pipeline~\cite{ai2022common} combining Monte-Carlo event simulation, detector response, and TrackML-style~\cite{kiehn-trackmlscore} postprocessing. \footnote{The full dataset pipeline is available at \href{https://github.com/EugeneShalli/hits-gen}{Hits-Gen}. The dataset for pileup 0 is available at \href{https://drive.usercontent.google.com/download?id=1-QQc4JyVSknCauGUsZFJQQbGEJR5rgMX&authuser=0}{Dataset Pileup 0}.}

We generate two processes: $pp \rightarrow t\bar{t}H,\, H\rightarrow b\bar{b}$ and inclusive $pp \rightarrow t\bar{t}$. Both are generated with \texttt{Pythia8}, producing stable truth-level particles as a starting point. Events are subsequently transported through a TrackML detector using the ACTS fast simulation (\texttt{Fatras}) and digitized into realistic measurements. This provides low-level hit data for machine learning models. From these digitized hits we further derive TrackML-style per-event triplets (\texttt{hits.csv}, \texttt{particles.csv}, \texttt{truth.csv}) with global coordinates and physics-motivated per hit weighting according to the TrackML paper.

We generate datasets at pileup levels 0, 5, 20, 50, and 200, each with 40k events with a 50-50 split between two processes.

\subsection{Improved model design}
\label{sec:new_model_design}

\subsubsection{Masking and projection}
The quadratic scaling of attention with hit count renders naive transformers impractical for full pixel-detector HL-LHC events. We address this with a hybrid design that combines geometric projection, clustering, and FlexAttention to exploit tracking locality.

\begin{figure}[htbp]
    \centering
    \includegraphics[width={0.6\textwidth}]{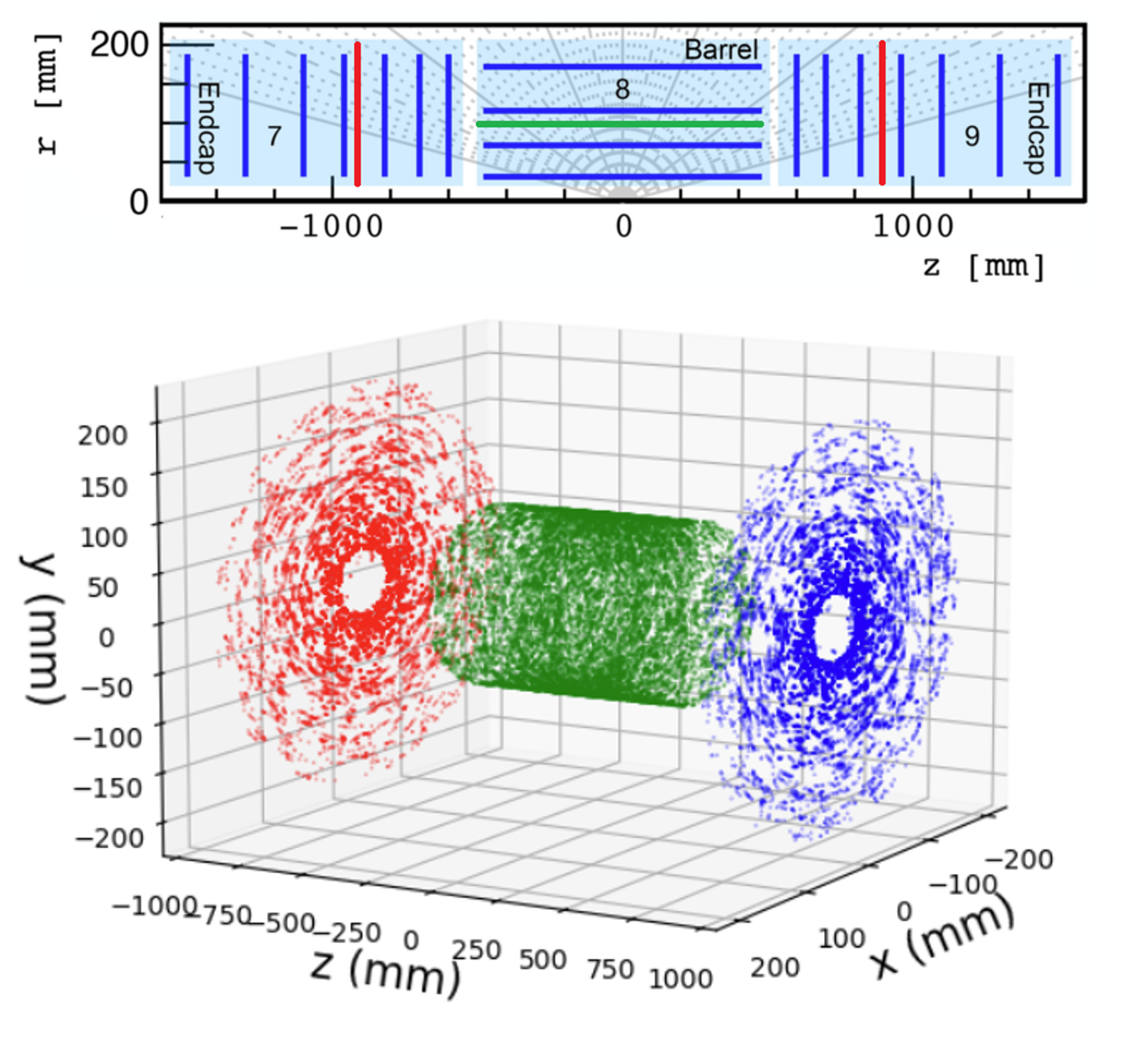}
    \caption{Top: the projection surfaces used for hit mapping, with the cylindrical barrel in green, two planar endcaps in red, and pixel detector layers in blue~\cite{kiehn-trackmlscore}. Bottom: pixel detector hits of one event projected onto these three surfaces.}
    \label{fig:hit_projection}
\end{figure}

As shown in \Cref{fig:hit_projection}, hits are projected onto simplified detector surfaces to minimize track spread: a cylinder ($R = 91\,\mathrm{mm}$) for the barrel (using $R$--$\phi$, $z$ coordinates) and two planes ($z = \pm 920\,\mathrm{mm}$) for the endcaps (using $x,y$). $R$ and $z$ are manually tuned hyperparameters. Tracks appear more compact and are more accurately recognized once projected onto the barrel surface. Tracks that are projected onto endcaps are nearly parallel to the beamline and appear as tight clusters, even though the true $z$-vertex position of the event may deviate from the point of origin. For these tracks, cluster alignment is refined by re-projecting clusters over candidate $z$-vertex positions and selecting the $z$ that maximizes alignment.

Light weight clustering (an iterative windowing algorithm developed by the authors) or DBSCAN~\cite{dbscan2022} is then applied on these projected surfaces to form local neighborhoods. Clusters on the cylindrical surface define sparse block masks for FlexAttention, ensuring that only physically plausible hit pairs attend and reducing the effective attention matrix by up to $\sim 400\times$. Endcap clusters use the vertex-$z$ scan described above to sharpen alignment in the longitudinal direction. Clustering hyperparameters are tuned to maximize the reconstructible ratio, defined as tracks with $p_T > 0.9\,\mathrm{GeV}$, having $\geq 3$ hits, and $\geq 50\%$ of those hits in a single cluster. Block masks are precomputed and cached for efficient reuse during training.

The encoder is a PyTorch~\cite{paszke2019pytorch} Transformer with FlexAttention (12 layers, 4 heads, hidden dimension $192$, feed-forward dimension $384$). Input $(x,y,z)$ hits are projected, clustered, and normalized. Training used AdamW with \texttt{bfloat16} mixed precision on NVIDIA H100s, gradient clipping, and an adaptive learning rate schedule. The model was trained on 8\,658 TrackML events with 96 validation and 96 test events.

Rather than regressing track parameters, the model maps each hit to a 32-dimensional embedding and is trained with a multi-positive InfoNCE~\cite{rusak2025infonce} contrastive loss: for each hit, all hits from the same track ($p_T > 0.9\,\mathrm{GeV}$) are positives, while all others are negatives. At inference, the model produces an $N \times N$ cosine-similarity matrix, from which tracks are assembled by selecting high-similarity neighbors, eliminating the need for a separate clustering stage.

\subsubsection{Joining regression and classification}
In this experiment, we combine our strongest prior architectures into a unified two-stage model. Stage~1 is an EncReg-style encoder-only Transformer~\cite{Caron_2025} that regresses track parameters $(\theta, \sin\phi, \cos\phi, q\!\in\!\{-1,1\})$ together with four learned free latent variables. Here $\theta$ and $\phi$ are spherical momentum angles ($p=\sqrt{p_x^2+p_y^2+p_z^2}$, 
$\theta=\arccos(p_z/p)$, $\phi=\arctan2(p_y,p_x)$). 

Stage~2 is an EncCla-style encoder-only Transformer for per-hit classification. For each hit we concatenate the raw coordinates with the regressor outputs, ($x$, $y$, $z$, $\theta$, $\sin\phi$, $\cos\phi$, $q$, $\text{latent}_1$, \ldots, $\text{latent}_4$), project to an embedding, and pass through encoder blocks. A final linear head produces a categorical distribution over quantile-binned $(\phi,\theta,p,q)$ classes; the predicted class is the maximum-probability bin. Although regressed parameters are already predictive, they further enrich the classifier’s input features.

The model is trained end-to-end with a joint loss $\mathcal{L} = \alpha \,\mathcal{L}_{\text{reg}} + \beta \,\mathcal{L}_{\text{cla}}$ ($\alpha=1$ and $\beta=0.3$), where $\mathcal{L}_{\text{reg}}$ is per-hit MSE on $(\theta,\sin\phi,\cos\phi,q)$ and $\mathcal{L}_{\text{cla}}$ is cross-entropy over class labels.  

The joint model, denoted JM $X$:$Y$ with $X$ regressor layers and $Y$ classifier layers, retains the one-pass property of both components: a single forward pass produces track parameters and per-hit classes, enabling downstream use without extra clustering stages.

\subsubsection{FlexAttention}
In our previous work~\cite{Caron_2025} we experimented with FlashAttention-2~\cite{dao_flashattention-2_2023}; here we instead adopt FlexAttention~\cite{dong_flex_2024}. This change is driven by a practical limitation of FlashAttention-2: while it supports variable sequence lengths, it requires packing sequences into a single concatenated tensor with manual offset tracking. As a result, training was restricted to a batch size of one to avoid manual batch-wise padding. FlexAttention overcomes this constraint through its BlockMask mechanism, which pre-computes tile-level sparsity, enabling efficient processing of heterogeneous sequence lengths within a standard batched tensor layout while maintaining near state-of-the-art kernel performance~\cite{dong_flex_2024}. With FlexAttention, we preserve the GPU inference speedups previously observed, now without the batch size restriction. Equally importantly, its memory efficiency allowed us to co-train both the regressor and classifier on a single NVIDIA A100 GPU (40\,GiB HBM2), whereas with FlashAttention only one of these models could fit on the same hardware during training.

\section{Results}
\label{sec:results}
Since the new datasets (Section~\ref{sec:new_datasets}) were developed concurrently with improved model design (Section~\ref{sec:new_model_design}), and also for easier comparison, the results below are based on a curated TrackML dataset~\cite{kiehn-trackmlscore} that is consistent with our prior work~\cite{Caron_2025}. This dataset has 200-500 tracks per event.
\subsection{Masking and projection}
Inference latency for the masking and projection pipeline can be broken down to: 6\,ms per event for clustering with parallel DBSCAN on projected surfaces, 2\,ms for block-mask creation, 20\,ms for the Transformer encoder, and 47\,ms for the track--hit assignment.

The resulting end-to-end runtime is on the order of tens of milliseconds per event, significantly faster than existing GNN pipelines (0.5--1\,s)~\cite{atlas2024gnn2} and comparable to the state of the art ($\sim$100\,ms)~\cite{vanstroud2024transformers3}.

In terms of physics performance, our model achieves $\sim$90\% track double-majority efficiency in the barrel and 91\% in the endcaps after vertex-$z$ refinement. Efficiencies in the barrel are uniform across $p_T$, with expected drops at low $|\eta|$ 
and near $|\eta|\!\approx\!2.0$ due to detector geometry.  

Relative to the EncReg and EncCla models from our previous iteration, which achieved $\sim$70\% efficiency on reduced TrackML datasets ($\sim$5\% HL-LHC density), the present design scales to tens of thousands of hits per event while below the 200\,ms inference latency in our previous work~\cite{Caron_2025} and improved efficiency. These results establish projection-based clustering with FlexAttention and contrastive similarity learning as a practical solution for HL-LHC scale tracking.

\subsection{Joining regression and classification}
The accuracy and TrackML score~\cite{kiehn-trackmlscore} for JM and EncCla models with FlexAttention are shown in \Cref{tab:scoretable}. Deeper models (more encoder layers) consistently improve both metrics. Adding EncReg and passing its regressed parameters to EncCla yields an additional $\sim$2.4\% accuracy and $\sim$2\% TrackML score gain. Unlike EncCla, the regressor showed little benefit from greater depth, so EncReg was kept shallow. 
\begin{table}[htbp]
    \centering
    \renewcommand{\arraystretch}{1.2}
    \setlength{\tabcolsep}{1pt}
    \begin{tabular}{lcccccccc}
        \toprule
        Model &
        \shortstack{EncCla\\6} &
        \shortstack{JM\\6:6} &
        \shortstack{EncCla\\7} &
        \shortstack{JM\\7:7} &
        \shortstack{EncCla\\9} &
        \shortstack{JM\\7:9} &
        \shortstack{EncCla\\15} &
        \shortstack{JM\\9:15} \\
        \midrule
        Accuracy   & 69.7\% & 72.8\% & 74.2\% & 76.6\% & 76.2\% & 78.4\% & \underline{78.5\%} & \textbf{80.5\%} \\
        TrackML score & 79.9\% & 82.3\% & 84.2\% & 86.4\% & 87.3\% & 89.0\% & \underline{89.8\%} & \textbf{91.4\%} \\
        \bottomrule
    \end{tabular}
    \caption{The accuracy and TrackML scores of different models across Joined Model (JM) and EncCla configurations. The best scores are print in bold and the second best are underlined. The numbers after the model name signal the layer depth. For JM configurations the first number denotes the layer depth of the regressor and the second number denotes the layer depth of the classifier. }
    \label{tab:scoretable}
\end{table}

Inference times are modest: CPU latency is stable at 0.1\,ms across all models, while GPU latency scales linearly with depth, adding $\sim$2.4\,ms per encoder layer on an NVIDIA A100 (40\,GiB).
\begin{table}[htbp]
    \centering
    \renewcommand{\arraystretch}{1.2}
    \setlength{\tabcolsep}{1pt}
    \begin{tabular}{lcccccccc}
        \toprule
        Model &
        \shortstack{EncCla\\6} &
        \shortstack{JM\\6:6} &
        \shortstack{EncCla\\7} &
        \shortstack{JM\\7:7} &
        \shortstack{EncCla\\9} &
        \shortstack{JM\\7:9} &
        \shortstack{EncCla\\15} &
        \shortstack{JM\\9:15} \\
        \midrule
        CPU inf. time (ms) & 0.1 & 0.1 & 0.1 & 0.1 & 0.1 & 0.1 & 0.1 & 0.1 \\
        GPU inf. time (ms) & 16.1 & 31.6 & 18.5 & 38.2 & 24.0 & 42.8 & 39.0 & 61.6 \\
        \bottomrule
    \end{tabular}
    \caption{The CPU inference time and GPU inference time per event in milliseconds of different models across Joined Model (JM) and EncCla configurations. The numbers after the model name signal the layer depth. For JM configurations the first number denotes the layer depth of the regressor and the second number denotes the layer depth of the classifier.}
    \label{tab:timetable}
\end{table}

Overall, scaling encoder-only Transformer trackers and coupling regression with classification in a single forward pass substantially improves performance over previous work. EncCla models show monotonic gains in TrackML score with depth, reaching 89\% (vs.\ 78\% previously). Injecting physics-based features from the regressor into the classifier provides a further $\sim$2\% absolute uplift. These improvements are enabled by FlexAttention, which allows deeper architectures to train on the same hardware, albeit with roughly doubled GPU inference time.  

\section{Conclusion and future work}
\label{sec:conclusion}
We release a fully reproducible ACTS-based hit-level dataset of $pp \rightarrow t\bar{t}H,\, H\rightarrow b\bar{b}$ signal and $pp \rightarrow t\bar{t}$ background events, providing TrackML-style formats across multiple pileup conditions (0--200) to enable realistic large-scale tracking benchmarks for machine learning models. In our experiments with new model architectures, we have shown that projection-based clustering combined with FlexAttention block masking provides an efficient way to scale transformer-based trackers to HL-LHC hit densities, cutting attention cost by up to $400\times$ while retaining end-to-end inference times in the $\mathcal{O}(10^2)$ ms range. In addition, deeper encoder-only architectures continue to deliver strong performance, while fusing regressed parameters into the classifier provides modest but consistent improvements. All of these gains are achieved within a single end-to-end inference call, preserving the simplicity that makes encoder-only designs appealing for HL-LHC deployment.

\section*{Acknowledgments}
This work used the Dutch national e-infrastructure with the support of the SURF Cooperative using grant no. EINF-11730. The work of R. RdA was supported by PID2020-113644GB-I00 from the Spanish Ministerio de Ciencia e Innovación and by the PROMETEO/2022/69 from the Spanish GVA. The author(s) gratefully acknowledges the computer resources at Artemisa, funded by the European Union ERDF and Comunitat Valenciana as well as the technical support provided by the Instituto de Fisica Corpuscular, IFIC (CSIC-UV).
